\documentclass[aps,prd,twocolumn,showpacs,floatfix,preprintnumbers,amsmath,amssymb,nofootinbib]{revtex4-1}

\input epsf
\usepackage{graphicx}
\usepackage{color}

\newcommand{\beq}{\begin{equation}}
\newcommand{\eeq}{\end{equation}}
\newcommand{\barr}{\begin{eqnarray}}
\newcommand{\earr}{\end{eqnarray}}

\newcommand{\rme}{\textrm{e}}

\newcommand{\Ly}{\textrm{Ly}}

\newcommand{\Tm}{T_{\rm m}}
\newcommand{\Tr}{T_{\rm r}}

\newcommand{\lsim}{\mathrel{\hbox{\rlap{\lower.55ex\hbox{$\sim$}} \kern-.3em \raise.4ex \hbox{$<$}}}}
\newcommand{\gsim}{\mathrel{\hbox{\rlap{\lower.55ex\hbox{$\sim$}} \kern-.3em \raise.4ex \hbox{$>$}}}}
\begin{document}
\title{Metals at the surface of last scatter}

\author{Yacine Ali-Ha\"imoud} 
\author{Christopher M.~Hirata} 
\author{Marc Kamionkowski}
\affiliation{California Institute of Technology, Mail Code 350-17, Pasadena, CA 91125}

\date{\today}
\begin{abstract}
Standard big-bang nucleosynthesis (BBN) predicts only a trace
abundance of lithium and no heavier elements, but some
alternatives predict a nonzero primordial 
metallicity.  Here we explore whether CMB measurements may set
useful constraints to the primordial metallicity and/or whether
the standard CMB calculations are robust, within the tolerance
of forthcoming CMB maps, to the possibility of primordial
metals.  Metals would affect the recombination history (and thus
CMB power spectra) in three ways:  (1) Ly$\alpha$ photons can be
removed (and recombination thus accelerated) by photoionizing
metals.  (2) The Bowen resonance-fluorescence mechanism may
degrade Ly$\beta$ photons and thus enhance the Ly$\beta$ escape probability and speed up recombination.  (3) Metals could affect the low-redshift tail
of the CMB visibility function by providing additional free
electrons.  The last two of these provide the strongest CMB signal.
However, the effects are detectable in the \emph{Planck} satellite
only if the primordial metal abundance is at least a few hundredths
of solar for (2) and a few tenths of solar for (3). We thus conclude that \emph{Planck} will not be able
to improve upon current constraints to primordial metallicity, at the
level of a thousandth of solar, from the Lyman-$\alpha$ forest and ultra-metal-poor halo stars,
and that the CMB power-spectrum predictions for \emph{Planck}
suffer no uncertainty arising from the possibility that there
may be primordial metals.
\end{abstract}

\pacs{98.80.Ft, 26.35.+c, 95.30.Jx}

\maketitle

\section{Introduction}
\label{sec:intro}

Big-bang nucleosynthesis (BBN) is one of
the pillars of the hot standard cosmological model. Comparison
of BBN theoretical predictions to observed abundances of the
lightest nuclei (D, $^3$He, $^4$He and $^7$Li) uniquely
determines the only free parameter of standard BBN, the
baryon-to-photon ratio $\eta = (5.7\pm 0.3) \times 10^{-10}$, or
equivalently, given the cosmic microwave background (CMB) temperature today $T_0 = 2.73$ K, the baryon abundance $\Omega_b h^2 = 0.021 \pm
0.001$ (see, e.g., Ref.~\cite{Iocco:2008va}). The latest
results from CMB anisotropy
measurements by the WMAP satellite are in excellent agreement,
with $\Omega_b h^2 = 0.02249^{+0.00056}_{-0.00057}$
\cite{Komatsu:2010fb}. 

In the standard BBN scenario, elements heavier than lithium are
only produced with trace abundances \cite{Iocco:2007km}. It has been shown, however,
that significant amounts of heavy elements may be produced in
inhomogeneous BBN (IBBN) models
\cite{Jedamzik:1994de,Matsuura:2005rb,Matsuura:2007sb}. IBBN may
take place if
some non-standard mechanism leads to large baryon-abundance
inhomogeneities on small scales, which are allowed by current
observations. It is possible to adjust the IBBN parameters to
reproduce the observed abundances of light elements, while
producing heavier elements with abundances as large as those in
the Sun \cite{Matsuura:2007sb,Nakamura:2010is}. More
generally, it cannot be excluded that some unknown processes may
lead to a significant production of metals heavier than
lithium.  It may therefore be profitable to inquire what
empirical constraints to primordial metals might be possible.

Standard methods to constrain metal abundances at high redshifts
rely on line emission or absorption measurements, and therefore
require some sources to have already formed (typically,
high-redshift quasars). The intergalactic medium (IGM) is
therefore already partially enriched by metals produced in the
first stars, and extracting limits on the primordial abundances
relies on understanding the complex physics of galactic outflows
and gas mixing and correctly modeling the ambient radiation
field. It would be of great interest to be able to probe the
abundance of metals \emph{before} the formation of Population
III stars that enriched the IGM. A few ideas were put forward to
probe the metallicity during the dark ages: Ref.~\cite{Loeb:2001ax} suggested using
resonant scattering of CMB photons off neutral lithium atoms (later shown to be unobservable because
lithium is kept ionized by redshifted Lyman-$\alpha$ photons
emitted during primordial hydrogen recombination
\cite{Switzer:2005nd}); Ref.~\cite{Basu:2004} studied the effect of fine-structure transitions of heavy elements in atomic or ionized states on CMB anisotropies; Ref.~\cite{Harwit:2003in} considered the spectral signatures of carbon and
oxygen. In this paper we assess whether
heavy elements present during primordial recombination could be
detectable from upcoming CMB experiments.

Primordial recombination has recently been the subject of a
renewed interest, due to the impact of uncertainties in the
standard theory on the predicted CMB temperature and
polarization anisotropy power spectrum. Errors in the
free-electron fraction $x_e(z)$ as small as a few tenths of a
percent near the peak of the visibility function at redshifts $z
\sim 1100$ would
induce biases of several standard deviations for cosmological
parameters estimated from \emph{Planck} data
\cite{Planck,RubinoMartin:2009ry}. This accuracy requirement has
motivated abundant
work on radiative transfer in the Lyman lines, in particular
Lyman-$\alpha$ (see for example
Refs.~\cite{Hirata:2008ny,Chluba:2009us,Hirata:2009qy,AliHaimoud:2010ym} and
references therein). The
tails of the visibility function are less important, but an
accuracy of about a percent is still needed, which required
implementing an accurate multi-level--atom formulation of the
recombination problem
\cite{Chluba:2006bc,Grin:2009ik,Chluba:2010fy,AliHaimoud:2010ab}. Such
a high sensitivity to 
the recombination history can be turned into an asset and serve
to probe unusual physics taking place during the recombination
history as, for example, the presence of primordial heavy
elements. In this paper, we explore this idea, and quantify the
impact of neutral metals on the Ly$\alpha$ and Ly$\beta$ net decay rates, and
of ionized metals on the low-redshift tail of the visibility
function.

Below we consider three effects of metals on the recombination
history and thus on the CMB visibility function:  (1) The removal of
Lyman-$\alpha$ photons (and thus acceleration of recombination)
by photoionization of metals (Section~\ref{sec:lymanalpha}); (2)
the degradation of
Lyman-$\beta$ photons (and thus acceleration of recombination) by the Bowen resonance-fluorescence mechanism
(Section~\ref{sec:bowen}); and (3)
the contribution to the free-electron abundance at late times by
low-ionization metals (Section \ref{sec:freeelectrons}).  We
find that effects (2) and (3) provide the biggest impact on CMB power spectra.
However, the effects are visible in \emph{Planck} only if the
primordial metal abundance is at least a few hundredths of solar for (2) and a few tenths of solar for (3).
Given that the Lyman-alpha forest \cite{Wolfe:2005} and ultra-metal-poor halo stars \cite{Beers:2005} constrain the primordial
metal abundance to be at least a few orders of magnitude
smaller than solar, we conclude that \emph{Planck} will be
unable to improve upon current constraints to the primordial
metal abundance or, alternatively, that the standard CMB
predictions for \emph{Planck} are robust to primordial metals at
the levels allowed by current empirical constraints.

\section{Effect of neutral metals on the Lyman-$\alpha$ decay
rate}
\label{sec:lymanalpha}

All metals (in the proper chemical sense of the term, i.e. not
including noble gases, halogens and other nonmetals) have a
first ionization energy below 10.2 eV, which corresponds 
to the Ly$\alpha$ transition in hydrogen. This means that
neutral metals can provide continuum opacity in the vicinity of
the Ly$\alpha$ line by absorbing Ly$\alpha$ photons in
photoionization events. Since the photoejected electrons rapidly
thermalize their energy, this results in a net loss of resonant
Ly$\alpha$ photons, which would have otherwise been reabsorbed
by ground state hydrogen atoms. The presence of metals can
therefore speed up hydrogen recombination by increasing the net
rate of Lyman-$\alpha$ decays. A similar process was
investigated for primordial helium recombination
\cite{Kholupenko:2007, Switzer:2007sn}: in that case the presence of neutral
hydrogen leads to continuum opacity in the He I $2^1P^o - 1^1S$
line. To estimate the impact of continuum opacity on the
Lyman-$\alpha$ line, we use the analytic treatment presented for
He I recombination in Ref.~\cite{AliHaimoud:2010dx}.

\subsection{Continuum opacity in Ly$\alpha$ due to
photoionization of neutral metals} 

The radiative-transfer equation in the vicinity of Ly$\alpha$
for the photon occupation number $f_{\nu}$, including only true
absorptions and emissions (i.e. neglecting resonant scatterings)
and continuum opacity is
\barr
\frac{1}{H \nu_{\rm Ly \alpha}}\frac{\partial f_{\nu}}{\partial t} - \frac{\partial f_{\nu}}{\partial \nu} &=& \tau_{\rm abs} \phi(\nu) \left(\frac{x_{2p}}{3 x_{1s}} - f_{\nu}\right) \label{eq:rad-trans}\nonumber\\
 &+& \eta_c \left(\rme^{- h \nu/\Tm} - f_{\nu}\right),
\earr
where we approximated $\nu \approx \nu_{\rm Ly \alpha}$ in the prefactor on the left-hand-side, $\tau_{\rm abs}$ is the Sobolev optical depth for true
absorption in the Ly$\alpha$ line, $\phi(\nu)$ is the line
profile, and $\eta_c$ is the continuum differential optical
depth, given by
\beq
\eta_c \equiv \frac{n_{\rm M^0} c \sigma_{\rm pi}(\nu_{\rm Ly \alpha})}{H \nu_{\Ly \alpha}}, \label{eq:etac}
\eeq 
In Eq.~(\ref{eq:etac}), $n_{\rm M^0}$ is the abundance of
neutral metal M$^0$, and $\sigma_{\rm pi}(\nu)$ is the
photoionization cross section of M$^0$ at frequency $\nu$. We have assumed that $\sigma_{\rm pi}$ varies slowly over the Ly$\alpha$ resonance (specifically, over the region which is optically thick for true absorption, which corresponds to a few tens of Doppler widths \cite{Hirata:2008ny}), so we can approximate $\sigma_{\rm pi}(\nu) \approx \sigma_{\rm pi}(\nu_{\rm Ly \alpha})$. Note that Eq.~(\ref{eq:rad-trans}) assumes that the ionization state
of M$^0$ is given by the Saha equilibrium equation (this translates in a ratio of continuum emission to absorption rates equal to $\rme^{-h\nu/\Tm}/f_{\nu}$), even though this
is not strictly correct (see Sec.~\ref{sec:ionization state}).

The net rate of $2p\rightarrow 1s$ decays is then obtained as follows:
\beq
\dot{x}_{2p\rightarrow 1s} = \frac{8 \pi \nu_{\Ly \alpha}^2}{c^3 n_{\rm H}} \int H \nu \tau_{\rm abs}\phi(\nu) \left(\frac{x_{2p}}{3 x_{1s}} - f_{\nu}\right) d \nu,
\eeq
where the prefactor converts photon occupation numbers to
photons per unit frequency per hydrogen atom, and we have
approximated $\nu \approx \nu_{\Ly \alpha}$ in the
multiplicative factor. Ref.~\cite{AliHaimoud:2010dx} showed that the net
decay rate in Lyman-$\alpha$ can be written in the following
form:
\beq
\dot{x}_{2p\rightarrow 1s} = \mathcal{E}\times \dot{x}_{2p\rightarrow 1s} \big{|}_{\rm std},
\eeq
where 
\beq
\dot{x}_{2p\rightarrow 1s} \big{|}_{\rm std} = \frac{8 \pi H \nu_{\Ly \alpha}^3}{c^3 n_{\rm H}} \left(\frac{x_{2p}}{3 x_{1s}} -\rme^{- h \nu_{\Ly \alpha}/\Tr}\right) 
\eeq
is the standard net decay rate in Ly$\alpha$, in the Sobolev
approximation, for a large optical depth and assuming an
incoming blackbody radiation field, and $\mathcal{E}$ is a
correction factor accounting for continuum absorption in the
line. The correction factor $\mathcal{E}(\tau_c)$ depends on the
single parameter
\beq
\tau_c \equiv \frac{\tau_{\rm abs} \Gamma_{2p} \eta_c}{4 \pi^2},\label{eq:tauc}
\eeq
where $\Gamma_{2p}$ is the total inverse lifetime of the $2p$
state. The dimensionless parameter $\tau_c$ can be interpreted
as the continuum optical depth within the part of the Ly$\alpha$
line which is optically thick for true absorption. For $\tau_c
\rightarrow 0$, $\mathcal{E}(\tau_c) \rightarrow 1$, and for
$\tau_c > 0$, $\mathcal{E}(\tau_c) > 1$, which is what one would
expect as continuum opacity increases the net rate of decays in
the line, as explained above. For $\tau_c \ll 1$, we have the
following approximate expansion (see Eq.~(117) of
Ref.~\cite{AliHaimoud:2010dx}):
\beq
\mathcal{E} \approx 1 + 13 \tau_c \ \ ,  \ \ \tau_c \ll 1.
\eeq 
In order for primordial metals to change the net decay rate in
Ly$\alpha$ by $\sim 1\%$ (which is roughly the level detectable
by \emph{Planck}), we therefore need $\tau_c \sim 0.001$ near
the peak of the visibility function. Extracting the relevant
parameters from the multilevel atom code of Hirata
\cite{Hirata:2008ny}, we obtain, for $z = 1100$,
\beq 
\tau_c \approx 0.7 \times 10^{-3} ~\frac{\sigma_{\rm pi}(\nu_{\rm Ly \alpha})}{10^{-17} \textrm{cm}^2} ~\frac{x_{\rm M^0}}{10^{-9}},
\eeq 
where $x_{\rm M^0} = n_{\rm M^0}/n_{\rm H}$ is the abundance of
neutral metals relative to hydrogen. We see that for a
characteristic photoionization cross section $\sigma_{\rm pi} =
10^{-17}$ cm$^2$, a fractional abundance of neutral metals per
hydrogen atom as low as $\sim 10^{-9}$ would be potentially
detectable.

\subsection{Ionization state of metals and results} \label{sec:ionization state}

We now turn to the evaluation of the fraction of neutral metals
$f_{\rm M^0} \equiv n_{\rm M^0}/n_{\rm M}$. As a first
approximation we use the Saha equilibrium value:
\beq
\frac{\left(1 - f_{\rm M^0}\right)}{f_{\rm M^0}}\Big{|}_{\rm Saha} = S_{\rm M} \equiv  \frac{g_{\rm M^+}g_e}{g_{\rm M^0}} \frac{\left(2 \pi m_e \Tr\right)^{3/2}}{n_e h^3} \rme^{- \chi_{\rm M}/\Tr}, \label{eq:Saha}
\eeq
where $\chi_{\rm M}$ is the ionization energy of M$^0$, $n_e$ is
the free electron abundance, and the $g$'s are the degeneracy
factors for each species. For a standard recombination history,
at $z = 1100$, Eq.~(\ref{eq:Saha}) gives (taking the ratio of
degeneracy factors to be unity) $f_{\rm M^0} = 5\times 10^{-3},
10^{-4},  2 \times 10^{-6}$ and $5 \times 10^{-8}$ for
$\chi_{\rm M} =$ 10, 9, 8 and 7 eV, respectively, and we can
anticipate that only metals with $\chi_{\rm M} \gtrsim 8$ eV may
have some impact on Ly$\alpha$.

Saha equilibrium assumes that the ionizing radiation field is
thermal. During hydrogen recombination, the radiation field
develops large distortions in the vicinity of the Ly$\alpha$
line, due to the slow escape of Ly$\alpha$ photons (in fact, thermalization of these distorsions is so inefficient that they survive until today \cite{Peebles}). These
non-thermal photons increase the ionization rate with respect to
its thermal value, and the neutral fraction of metals is
therefore smaller than predicted by the Saha equation (see
for example Ref.~\cite{Switzer:2005nd} for the case of
lithium). The ionization state of the metal M is therefore
rather determined by the balance of recombinations and
photoionizations (this assumes the steady-state limit, valid so
long as the photoionization rate is much larger than the Hubble
expansion rate, which is a very good approximation around the
peak of the visibility function)
\beq
n_{\rm M^+} n_e \alpha_{\rm M} = n_{\rm M^0} \beta_{\rm M}, \label{eq:M0M+ balance}
\eeq
where $\alpha_{\rm M}$ is the M$^+ \rightarrow $M$^0$
recombination coefficient and $\beta_{\rm M}$ is the M$^0
\rightarrow $M$^+$ photoionization rate. From Eq.~(\ref{eq:M0M+
balance}), we obtain the neutral fraction,
\beq
f_{\rm M^0} = \left(1 + \frac{\beta_{\rm M}}{n_e \alpha_{\rm M}}\right)^{-1}. \label{eq:fM0}
\eeq
The photoionization rate $\beta_{\rm M} = \beta_{\rm CMB} +
\beta_{\rm dist}$ comprises a thermal part, due to
photoionizations by CMB photons (from the ground state
\emph{and} excited states), which is related to the
recombination coefficient through the detailed balance relation
\beq
\beta_{\rm CMB} =  S_{\rm M} n_e \alpha_{\rm M},
\eeq
and a non-thermal part, due to photoionizations from the ground
state by distortion photons,
\beq
\beta_{\rm dist} = \int_{\nu_{\rm M}}^{\infty} \sigma_{\rm pi}(\nu) \frac{8 \pi \nu^2}{c^3} \Delta f_{\nu} \ d \nu, \label{eq:betaM}
\eeq
where $\nu_{\rm M} \equiv \chi_{\rm M}/h$ and $\Delta f_{\nu} =
f_{\nu} - f_{\nu}^{(\rm CMB)}$ is the non-thermal distortion to
the photon occupation number. The Lyman-$\alpha$ distortion
peaks around $z \sim 1400$.  We can therefore expect that
distortions may start significantly affecting the ionization
state of the metal M around redshift $z \sim 1400 ~ \chi_{\rm
M}/(10.2 ~\rm eV)$. As a consequence, we expect the Saha
equilibrium approximation to be quite accurate around $z \sim
1100$ for metals with ionization threshold lower than $\sim 8$
eV. For the more interesting metals with $\chi_{\rm M}\gtrsim 8$
eV, however, spectral distortions will lower the neutral
fraction with respect to the Saha value at $z \sim 1100$, making
their detection more difficult through the effect considered here (we will consider the effect of additional free electrons due to the presence of ionized metals in Sec.~\ref{sec:freeelectrons}).

\begin{figure} 
\includegraphics[width = 85mm]{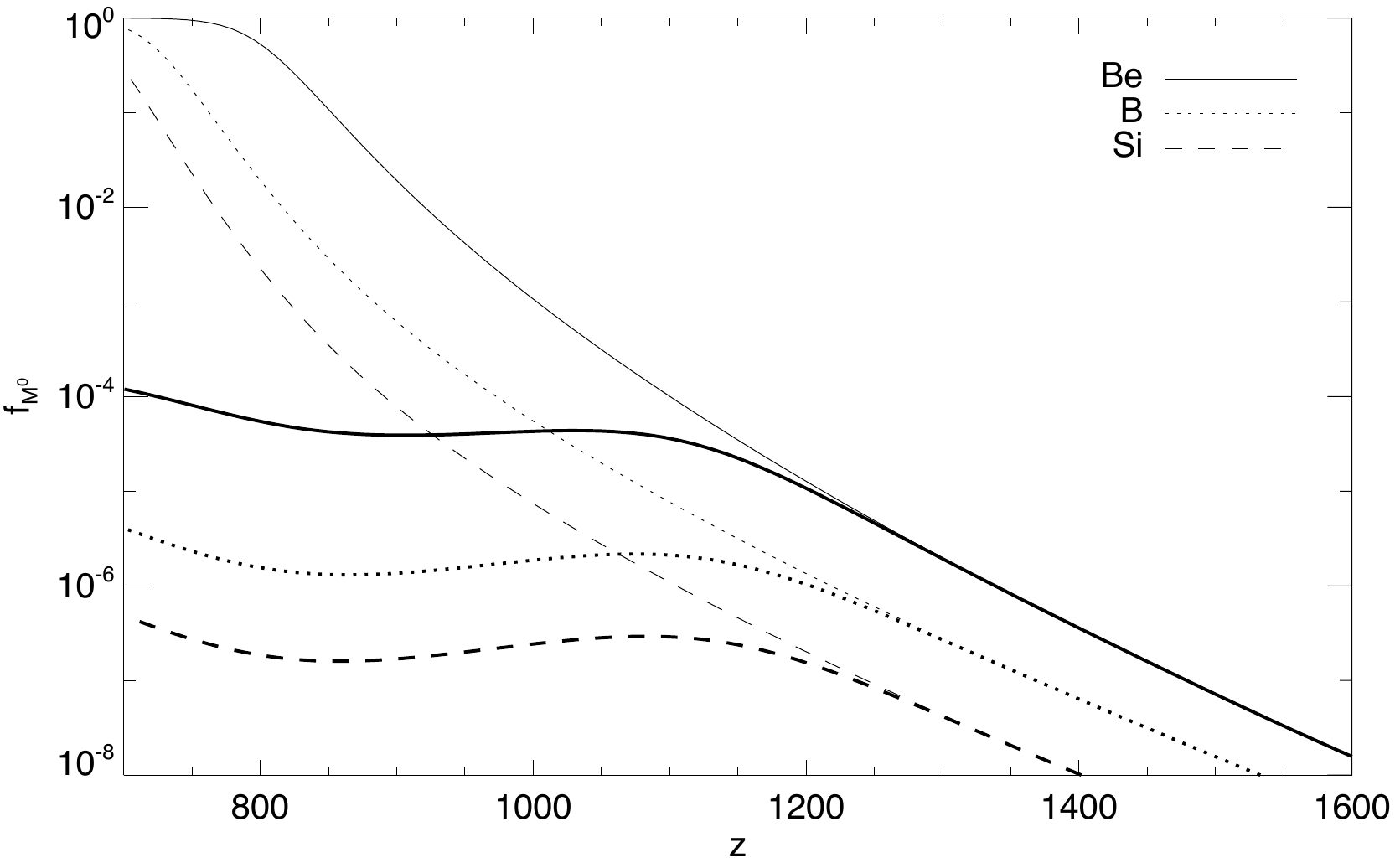}
\caption{Neutral fraction of beryllium, boron and silicon as a
function of redshift. Thin lines represent the Saha equilibrium
value given by Eq.~(\ref{eq:Saha}). Thick lines represent a more
accurate estimate accounting for distortions to the ambient
blackbody field near Lyman-$\alpha$.}
\label{fig:neutral_fraction}
\end{figure}

We have computed the neutral fraction of several metals with
atomic number $Z \leq 26$, using the fits of
Ref.~\cite{Verner:1996th} for the photoionization cross sections,
and the \textsc{chianti} database for the recombination
coefficients \cite{Chianti1,Chianti6}. We have extracted the
Ly$\alpha$ distortion\footnote{In principle, to be
self-consistent, one should account for the continuum optical
depth due to metal photoionization above Ly$\alpha$ and between
$\nu_{\rm M}$ and $\nu_{\rm Ly \alpha}$. Given that we find that
this effect should not be detectable anyway, we have not
implemented a more subtle treatment.} from the two-photon code
of Hirata \cite{Hirata:2008ny}. We show the ionization state of
beryllium ($\chi_{\rm Be} = 9.32$ eV), boron ($\chi_{\rm B} =
8.30$ eV) and silicon ($\chi_{\rm Si} = 8.15$ eV) as a function
of redshift, for a standard recombination history, in
Fig.~\ref{fig:neutral_fraction}.

\begin{figure} 
\includegraphics[width = 85mm]{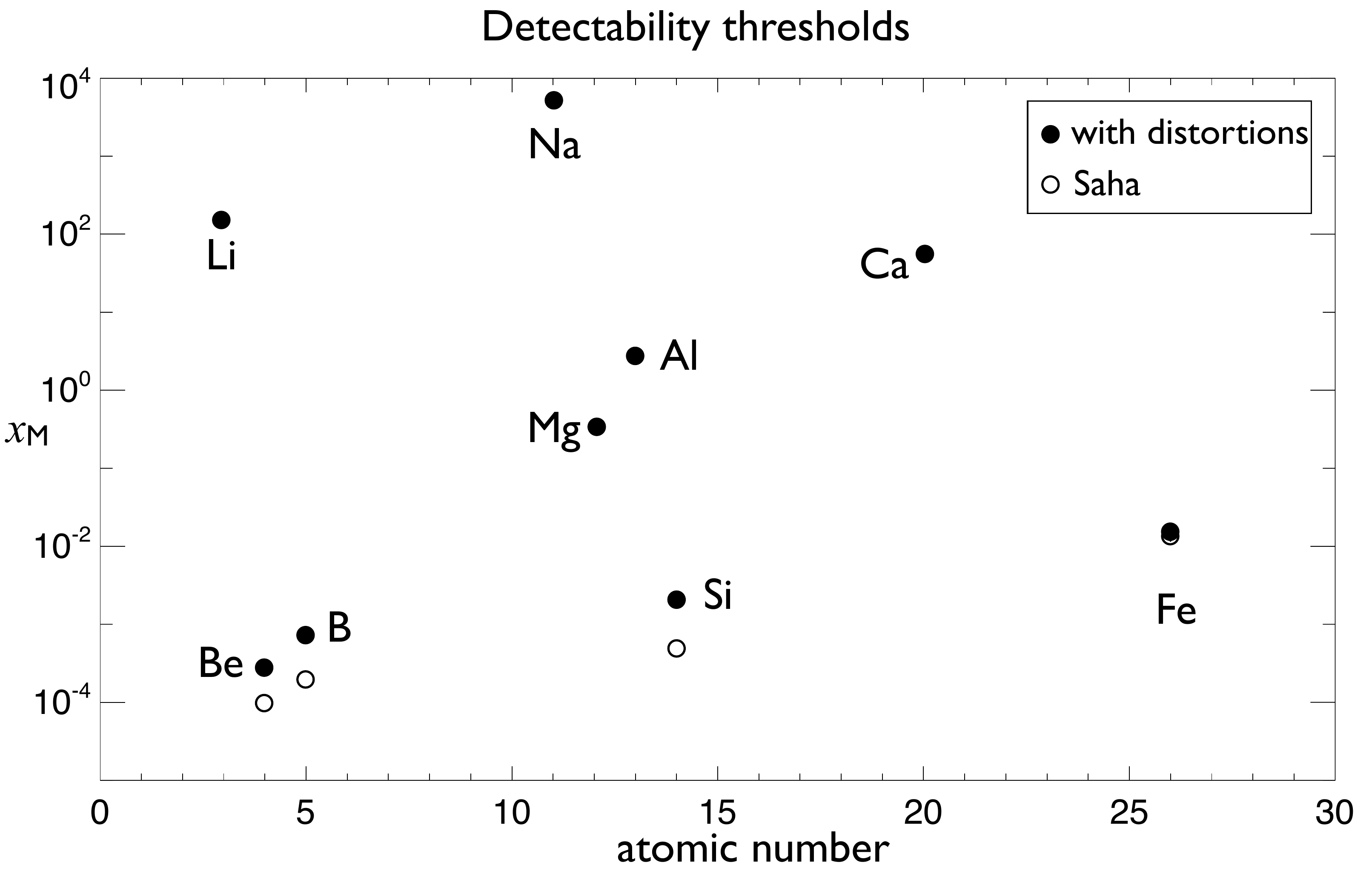}
\caption{Minimum abundance of metals relative to hydrogen needed
to result in a continuum optical depth in Ly$\alpha$ $\tau_c
\geq 0.001$ at $z = 1100$. }
\label{fig:threshold}
\end{figure}

We show in Fig.~\ref{fig:threshold} the minimal abundance of
metals detectable through its effect on Ly$\alpha$ (i.e. such
that $\tau_c \geq 0.001$ at redshift 1100). We see that the
smallest detectable abundance would be $x_{\rm Be} \sim 3\times
10^{-4}$. Due to lack of data, we have not treated the case of
other metals with $\chi_{\rm M} > 9$ eV, such as zinc
($\chi_{\rm Zn} = 9.39$), arsenic ($\chi_{\rm As} = 9.79$) and
gold ($\chi_{\rm Au} = 9.23$), but do not expect significantly
lower detectability thresholds unless they have unusually high
photoionization cross sections.

\section{ The Bowen resonance-fluorescence mechanism for
oxygen} 
\label{sec:bowen}

There is an accidental coincidence between the hydrogen
Ly$\beta$ line at 1025.72 \AA \ and the O I $2p^4~^3P_2 -
2p^33d~^3D^o_3$ line at 1025.76 \AA. Ly$\beta$ photons may
therefore excite neutral oxygen instead of being reabsorbed in
the hydrogen line. Neutral oxygen in the $2p^33d~^3D^o_3$ state
can then either directly decay back to the ground state or first
decay to the $2p^33p~^3P_2$ state by emitting an infrared photon
at 1.13 $\mu$m, and subsequently cascade down to the ground
state (in principle, atoms in the $2p^33d~^3D^o_3$ state can
also be excited or photoionized, but there are very few thermal
photons energetic enough to do so). The probability of the latter series of events (neglecting stimulated decays) is $p_{1.13 \mu \rm m} = A_{1.13\mu \rm m}/(A_{1.13\mu \rm m} + A_{1025.76}) \approx 0.3$. Direct decays of excited
oxygen to the ground state do not affect radiative transfer in
the Ly$\beta$ line, as they do not change the number of Ly$\beta$ photons. On the other hand,
absorptions in the 1025.76 \AA~line followed by emission of
infrared photons degrade Ly$\beta$ photons that would otherwise
have been reabsorbed by neutral hydrogen. This effect is similar
to the continuum opacity in Ly$\alpha$ discussed in Section
\ref{sec:lymanalpha},
except that this is now a resonant process. 

The escape\footnote{The term ``escape'' is somewhat misleading in this situation: photons near the Ly$\beta$ frequency do not actually escape more from the resonant region (in fact, their overall escape rate is even lower due to a slightly higher optical depth). They rather only ``escape'' reabsorption by neutral hydrogen.} probability in the Ly$\beta$ line is enhanced by the probability that a Ly$\beta$ photon is absorbed by O I (and then degraded) rather than by H I:
\beq
\Delta P_{\rm esc} \approx \frac{\frac{7}{9}n_{\textrm{OI}} A_{1025.76} p_{1.13 \mu \rm m}}{3 n_{\textrm{HI}} A_{3p,1s}} \approx 0.2  \frac{n_{\textrm{OI}}}{n_{\textrm{HI}}}, \label{eq:DPesc}
\eeq
where the multiplicative factors are the ratios of the degeneracy factors of the excited levels to those of the ground states.
Note that the photon occupation number redward of Ly$\beta$ is slightly decreased by this process: $f_- = x_{3p}/(3 x_{1s}) \left(1 - \Delta P_{\rm esc}\right)$. However, as long as $\Delta P_{\rm esc} \ll 1$ this has no detectable impact on radiative transfer redward of Ly$\beta$.

As oxygen and
hydrogen have very similar ionization energies, we can assume
that they have the same recombination history. More
specifically, the ratio of ionized to neutral oxygen rapidly
equilibrates to the corresponding ratio for hydrogen because the
continuum above the ionization threshold is optically thick. We therefore have $n_{\rm OI}/n_{\rm OII} \approx \frac98 n_{\rm HI}/n_{\rm HII}$ (the 9/8 comes from properly accouting for degeneracy factors, see Eq.~(\ref{eq:Saha})) and as a result we
obtain $n_{\rm OI}/n_{\rm HI} \approx x_{\rm O}/(1 - x_e/9) \approx x_{\rm O}$.
As long as $x_{\rm OI} \lesssim 10^{-3}$ the damping wings of the O I 1025.76 \AA~line are optically thin, and the Bowen mechanism can only affect the net decay rate in the Doppler core of the Ly$\beta$ line. The latter is very small anyway as the radiation field is very close to equilibrium with the 3p-1s ratio over many Doppler widths near line center. We have modified the escape probability from the Doppler core of Ly$\beta$ in the recombination code \textsc{HyRec} \cite{AliHaimoud:2010dx} according to Eq.~(\ref{eq:DPesc}), and found that a minimal abundance of oxygen $x_{\rm O} \approx 10^{-5}$ is required to affect the recombination history at a potentially detectable level $\Delta x_e/x_e \approx 0.2 \%$ at $z \approx 1100$. Note that this would correspond to an enhancement by a factor $\sim 200$ from the standard escape probability in the Ly$\beta$ Doppler core. This stems from the fact that only a tiny fraction of Ly$\beta$ decays ocurr in the Doppler core of the line, whereas most of them take place in the damping wings. The recombination history is therefore highly insensitive to the exact decay rate in the core.


\section{Additional free electrons due to ionized metals}
\label{sec:freeelectrons}

If metals remain ionized, they can contribute an additional
residual free-electron fraction at late times, $\Delta x_e \sim
x_{\rm M^+}$. In fact we have $\Delta x_e = \frac12 x_{\rm
M^+}$, as we show below. At late times the evolution of the free-electron
fraction is given by
\beq
\dot{x}_e \approx -n_{\rm H}\alpha_{\rm B} x_e x_p =  -n_{\rm H}\alpha_{\rm B} x_e (x_e - x_{\rm M^+}).\label{eq:xedot-late}
\eeq
Eq.~(\ref{eq:xedot-late}) is valid because the free electron
fraction is many orders of magnitude above the Saha equilibrium
value at late time (for a discussion, see Ref.~\cite{AliHaimoud:2010dx}).
If $x_e^0$ is the unperturbed free electron fraction
(i.e. obtained with $x_{\rm M^+} = 0$) and $x_e = x_e^0 + \Delta
x_e$, we obtain
\beq
\dot{\Delta x_e} = - n_{\rm H} \alpha_{\rm B}x_e^0\left(2 \Delta x_e - x_{\rm M^+}\right),
\eeq
which asymptotes to $\Delta x_e = \frac12 x_{\rm M^+}$.
The \emph{Planck} satellite will be sensitive to fractional
changes $\Delta x_e/x_e \sim 1\%$ at late times. Since $x_e
\approx 0.3 - 1 \times 10^{-3}$ for $200 \lesssim z \lesssim
700$, we conclude that a potential detection by \emph{Planck}
requires a fractional abundance of metals $x_{\rm M} \gtrsim
10^{-5}$ (in the case that metals remain fully ionized). Note
that for a given $\Omega_b$, the presence of metals also
modifies the total abundance of hydrogen, $n_{\rm H}$,
throughout the recombination history. However these
modifications are degenerate with a mere change of $\Omega_b$ of
$Y_{\rm He}$ at the level of a few times $10^{-5}$ and are
therefore undetectable.

\section{Conclusions} 

We have investigated whether a
primordial metal content could sufficiently affect the
recombination history to be detectable in upcoming CMB data from
\emph{Planck}. We first considered the effect of photoionization
of neutral metals by Ly$\alpha$ photons. We showed that although
a very small abundance of neutral metals would be enough to
significantly affect the net decay rate in Ly$\alpha$, metals
with ionization threshold below Ly$\alpha$ are mostly ionized at
$z\sim 1100$, and therefore undetectable.  We also considered
the Bowen resonance-fluorescence mechanism if primordial oxygen is present. This effect leads to an enhanced escape rate of Ly$\beta$ photons and a speed up of recombination. We showed that it could lead to detectable changes for a primordial oxygen abundance of a couple hundredths of solar $x_{\rm O} \sim 10^{-5}$.
Finally, we pointed out
that metals that stay ionized until late times provide
additional free electrons and therefore change the late-time
Thomson scattering optical depth. A fractional abundance $x_{\rm M}\sim 10^{-5}$
of primordial metals could be detectable through this effect. As a reference, the
most abundant metal in the solar photosphere is oxygen ($x_{\rm
O} = 4.9 \times 10^{-4}$), followed by carbon ($x_{\rm C} = 2.7
\times 10^{-4}$), neon ($x_{\rm Ne} = 8.5 \times 10^{-5}$),
nitrogen ($x_{\rm N} = 6.8 \times 10^{-5}$), magnesium ($x_{\rm
Mg} = 3.4 \times 10^{-5}$), silicon ($x_{\rm Si} = 3.2 \times
10^{-5}$), iron ($x_{\rm Fe} = 3.2 \times 10^{-5}$) and sulfur
($x_{\rm S} = 1.3 \times 10^{-5}$). Other metals have fractional
abundances $x_{\rm M} < 10^{-5}$ in the Sun \cite{Asplund:2009fu}. As
carbon, nitrogen, oxygen and neon are neutral at late times (due
to their high ionization potential), we conclude that
\emph{Planck} could potentially detect primordial metals with an
abundance at least a few tenths of solar. This is moreover an
optimistic estimate, as the effect of metals is likely to be
degenerate with the redshift of reionization or other
cosmological parameters.  

Given that Lyman-alpha-forest
measurements and ultra-metal-poor halo stars suggest a primordial metallicity much
smaller than one hundredth solar, we conclude that the CMB can
unfortunately not usefully constrain the abundance of primordial
metals.  At the same time, we also conclude that the CMB
predictions for the \emph{Planck} satellite are robust to a
primordial metallicity allowed by current empirical constraints.

\begin{acknowledgments} 
MK thanks the support of the Miller Institute for Basic
Research in Science at the University of California, Berkeley,
where part of this work was completed. This work was supported
by DoE DE-FG03-92-ER40701, NASA NNX10AD04G (MK), NSF AST-0807337
(YAH and CH), the Gordon and Betty Moore Foundation, the Alfred
P. Sloan Foundation (CH) and the David \& Lucile Packard
Foundation (CH).
\end{acknowledgments}


\begin{thebibliography}{99}



\bibitem{Iocco:2008va}
 F.~Iocco, G.~Mangano, G.~Miele, O.~Pisanti and P.~D.~Serpico,
 Phys.\ Rept.\  {\bf 472}, 1 (2009)

\bibitem{Komatsu:2010fb}
 E.~Komatsu {\it et al.}  [WMAP Collaboration],
Astrophys.\ J.\ Supp. {\bf 192}, 18 (2011)

\bibitem{Iocco:2007km}
 F.~Iocco, G.~Mangano, G.~Miele, O.~Pisanti and P.~D.~Serpico,
Phys.\ Rev.\  D {\bf 75}, 087304 (2007)
 [arXiv:astro-ph/0702090].

\bibitem{Jedamzik:1994de}
 K.~Jedamzik, G.~M.~Fuller, G.~J.~Mathews and T.~Kajino,
 Astrophys.\ J.\  {\bf 422}, 423 (1994)

\bibitem{Matsuura:2005rb}
 S.~Matsuura, S.~I.~Fujimoto, S.~Nishimura, M.~A.~Hashimoto and K.~Sato,
 Phys.\ Rev.\  D {\bf 72}, 123505 (2005)

\bibitem{Matsuura:2007sb}
  S.~Matsuura, S.~-i.~Fujimoto, M.~-a.~Hashimoto {\it et al.},
  Phys.\ Rev.\  {\bf D75}, 068302 (2007).

\bibitem{Nakamura:2010is}
 R.~Nakamura, M.~a.~Hashimoto, S.~i.~Fujimoto, N.~Nishimura and K.~Sato,
 arXiv:1007.0466 [astro-ph.CO].

\bibitem{Loeb:2001ax}
 A.~Loeb,
 Astrophys.\ J.\ Lett.\ {\bf 555}, L1 (2001)

\bibitem{Switzer:2005nd}
 E.~R.~Switzer and C.~M.~Hirata,
 Phys.\ Rev.\  D {\bf 72}, 083002 (2005)

\bibitem{Basu:2004}
K.~Basu, C.~Hern\'{a}ndez-Monteagudo and R.~A.~Sunyaev,
 Astron.\ Astrophys.\ {\bf 416}, 447 (2004) 

\bibitem{Harwit:2003in}
 M.~Harwit and M.~Spaans,
 Astrophys.\ J.\  {\bf 589}, 53 (2003)

\bibitem{Planck}
  The Planck Collaboration, 
  arXiv:astro-ph/0604069.

\bibitem{RubinoMartin:2009ry}
 J.~A.~Rubino-Martin, J.~Chluba, W.~A.~Fendt and B.~D.~Wandelt,
Mon.\ Not.\ Roy.\ Astron.\ Soc.\  {\bf 403}, 439 (2010)

\bibitem{Hirata:2008ny}
 C.~M.~Hirata,
 Phys.\ Rev.\  D {\bf 78}, 023001 (2008)

\bibitem{Chluba:2009us}
 J.~Chluba and R.~A.~Sunyaev,
 Astron.\ Astrophys.\ {\bf 512}, A53 (2010)

\bibitem{Hirata:2009qy}
 C.~M.~Hirata and J.~Forbes,
 Phys.\ Rev.\  D {\bf 80}, 023001 (2009)

\bibitem{AliHaimoud:2010ym}
 Y.~Ali-Haimoud, D.~Grin and C.~M.~Hirata,
 Phys.\ Rev.\  D {\bf 82}, 123502 (2010)

\bibitem{Chluba:2006bc}
 J.~Chluba, J.~A.~Rubino-Martin and R.~A.~Sunyaev,
 Mon.\ Not.\ Roy.\ Astron.\ Soc.\  {\bf 374}, 1310 (2007)

\bibitem{Grin:2009ik}
 D.~Grin and C.~M.~Hirata,
 Phys.\ Rev.\  D {\bf 81}, 083005 (2010)

\bibitem{Chluba:2010fy}
 J.~Chluba, G.~M.~Vasil and L.~J.~Dursi,
Mon.\ Not.\ Roy.\ Astron.\ Soc.\  {\bf 407}, 599 (2010)

\bibitem{AliHaimoud:2010ab}
 Y.~Ali-Haimoud and C.~M.~Hirata,
 Phys.\ Rev.\  D {\bf 82}, 063521 (2010)

\bibitem{Wolfe:2005}
A.~M.~Wolfe, E.~Gawiser and J.~X.~Prochaska, 
Ann.\ Rev.\ Astron.\ Astrophys.\ {\bf 43}, 861 (2005) 

\bibitem{Beers:2005}
T.~C.~Beers and N.~Christlieb, 
Ann.\ Rev.\ Astron.\ Astrophys.\ {\bf 43}, 531 (2005)

\bibitem{Kholupenko:2007}
E.~E.~Kholupenko, A.~V.~Ivanchik and D.~A.~Varshalovich,
Mon.\ Not.\ Roy.\ Astron.\ Soc.\  {\bf 378}, L39 (2007)

\bibitem{Switzer:2007sn}
 E.~R.~Switzer and C.~M.~Hirata,
 Phys.\ Rev.\  D {\bf 77}, 083006 (2008)

\bibitem{AliHaimoud:2010dx}
 Y.~Ali-Haimoud and C.~M.~Hirata,
 arXiv:1011.3758 [astro-ph.CO].

\bibitem{Peebles}
P.~J.~E.~Peebles, Astrophys.\ J.\  {\bf 153}, 1 (1968)

\bibitem{Verner:1996th}
 D.~A.~Verner, G.~J.~Ferland, K.~T.~Korista and D.~G.~Yakovlev,
 Astrophys.\ J.\  {\bf 465}, 487 (1996)

\bibitem{Chianti1}
  K.~P.~Dere {\it et al.}, 
  Astron.\ Astrophys.\ Supp.\ Ser.\ {\bf 125}, 149 (1997).

\bibitem{Chianti6}
  K.~P.~Dere {\it et al.}, 
  Astron.\ Astrophys.\ Supp.\ Ser.\ {\bf 498}, 915 (2009).

\bibitem{Asplund:2009fu}
 M.~Asplund, N.~Grevesse, A.~J.~Sauval and P.~Scott,
 Ann.\ Rev.\ Astron.\ Astrophys.\  {\bf 47}, 481 (2009)






\end{thebibliography}
\end{document}